\documentclass[aps,prb,reprint,showpacs,superscriptaddress]{revtex4-1}

\usepackage{graphicx}

\begin{document}

\title{Influence of the $A$ cation on the low-temperature antiferromagnetism of ordered antiferroelectric $A_2$CoTeO$_6$ perovskites }

\author{R. Mathieu}\email{roland.mathieu@angstrom.uu.se}
\affiliation{Department of Engineering Sciences, Uppsala University, Box 534, SE-751 21 Uppsala, Sweden}

\author{S. A. Ivanov}
\affiliation{Department of Inorganic Materials, Karpov' Institute of Physical Chemistry, Vorontsovo pole, 10 105064, Moscow K-64, Russia}

\author{R. Tellgren}
\affiliation{Department of Materials Chemistry, Uppsala University, Box 538, SE-751 21 Uppsala, Sweden}

\author{P. Nordblad}
\affiliation{Department of Engineering Sciences, Uppsala University, Box 534, SE-751 21 Uppsala, Sweden}

\date{\today}

\begin{abstract}

We have investigated the structural and magnetic properties of antiferroelectric $A_2$CoTeO$_6$ perovskites with $A$ = Cd, Ca, Sr, Pb, and Ba. All compounds are antiferromagnetic at low temperatures, with the antiferromagnetic transition temperature slightly decreasing with decreasing ionic size of the $A$ cation. Such a decrease in antiferromagnetic interaction is not observed in $A^{2+}_2$Co$M^{6+}$O$_6$ materials with other $M^{6+}$ cations of similar ionic sizes, suggesting that Te$^{6+}$ cations affect the electronic and in turn the magnetic structure of the system. 

\end{abstract}


\maketitle

\section{INTRODUCTION}

Decades after their discovery\cite{first}, multiferroics systems in which magnetic and electric polarizations are coupled are recently being (re-)investigated\cite{spaldin-fiebig, daniel, tokura-intro} due to their potential use in spintronic applications. In multiferroic-based devices, both polarizations could be used for storage or read/write operations\cite{fert}, or the electric polarization could be controlled by magnetic fields\cite{tokura,cheong}, or the magnetization could be altered using an electric field\cite{thomas}.

\begin{figure}
\includegraphics[width=0.46\textwidth]{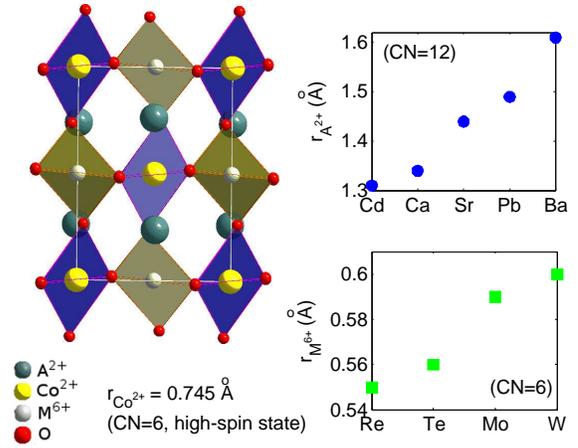}
\caption{(Color online) Illustration of the cation arrangement in an ordered monoclinic ($P2_1/n$) perovskite and ionic radii in {\AA} for different cations in $A_2$Co$M$O$_6$. CN refers to the coordination number; ionic radii extracted from Ref.~\onlinecite{shannon}}
\label{fig-intro}
\end{figure}

$A_2$$BB'$O$_6$ perovskites ($A$: alkaline rare-earth, $B$, $B'$: transition metals) are an interesting class of materials\cite{woodward}, with a relatively simple structure (see Fig.~\ref{fig-intro} for a schematic view). Interestingly these strongly correlated perovskites exhibit a wide variety of magnetic and electrical properties related to the coupling between lattice, orbital, and spin degrees of freedom. Most famous members of this family include the Sr$_2$FeMoO$_6$\cite{MoO} and Sr$_2$CrReO$_6$\cite{ReO} perovskites with high ferromagnetic Curie temperatures, or the ordered ferromagnetic La$_2$FeCrO$_6$\cite{LaFeCrO}. It was later predicted that the related Bi$_2$FeCrO$_6$ compound\cite{BiFeCrO} should exhibit multiferroic properties, owning to its structural properties and Fe/Cr cation ordering. Recently La$_2$NiMnO$_6$ was experimentally found to exhibit promising ferromagnetic and dielectric properties\cite{LNMO,ddUU}.

In this article, we review and analyze the structural and magnetic properties of antiferroelectric $A_2$CoTeO$_6$ perovskites with $A$ = Cd, Ca, Sr, Pb, and Ba.  In this system, the magnetic (Co$^{2+}$, $3d^{7}$) and diamagnetic (Te$^{6+}$, $4d^{10}$) cations are ordered on the ($B, B'$)-sites of the structure. The compounds with $A$ = Cd, Ca, Sr, and Pb adopt a monoclinic structure at low temperature, while in order to accommodate the larger Ba cation, Ba$_2$CoTeO$_6$ adopts a hexagonal structure\cite{bcto,pcto}. The average $<$Co-O$>$ bond length was found to vary only slightly, and in a non-monotonous fashion, with the ionic radius of the $A$ cation $r_A$. On the other hand the average $<$Co-O-Te$>$ bond angle was found to monotonously vary with $r_A$ from about 145 to 177 degrees. We observe that all compounds are antiferromagnetic at low temperatures (below 20 K), and that the N\'eel temperature (or antiferomagnetic transition temperature) $T_N$ slightly decreases with decreasing size of the $A$ cation. Such a decrease in $T_N$ is not observed in related $A_2$Co$M$O$_6$ materials with other $M^{6+}$ cations of similar sizes, suggesting that Te$^{6+}$ ions have an effect on the electronic structure of the materials.

\section{EXPERIMENTS}

High-quality polycrystalline $A_2$CoTeO$_6$ perovskites with $A$= Cd, Ca, Sr, Pb, and Ba were synthesized by conventional solid state reactions, as described in Refs.~\onlinecite{bcto,pcto}. The phase purity, and cation stoichiometry was checked by room-temperature x-ray diffraction (XRD) and energy-dispersive spectroscopy (EDS). Temperature-dependent XRD was performed on all samples, while temperature-dependent neutron powder diffraction (NPD) was performed on the samples with $A$ = Pb and Ba \cite{bcto,pcto}. Similar NPD measurements on compounds with $A$ = Ca and Sr \cite{alonso} have been reported by other authors, whose results have been included in our study for comparison. Cd$_2$CoTeO$_6$ could not be investigated using NPD, Cd being a very strong neutron absorber. The magnetization and heat capacity data of all samples but Sr$_2$CoTeO$_6$ were collected using a SQUID magnetometer (MPMSXL) and a Physical Properties Measurement System (PPMS6000) from Quantum Design Inc, respectively.

\begin{figure}[htb]
\includegraphics[width=0.46\textwidth]{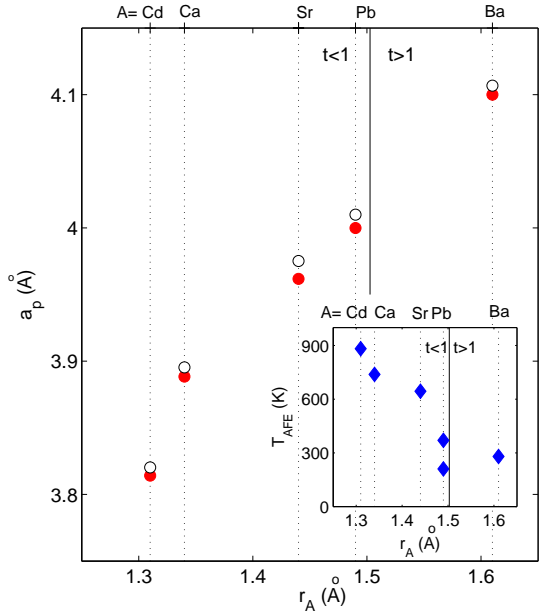}
\caption{(Color online) Dependence on the $A$-cation ionic radius $r_A$ of the pseudo-cubic unit cell parameter $a_p = (\frac{V}{Z})^{\frac{1}{3}}$ ($V$ is the cell volume, $Z$ = 4 or 6 formula units per cell). Open symbols denote room-temperature (295 K) data, while filled symbols indicate low-temperature ($<$ 10 K) data. The inset shows the variation of the antiferroelectric transition temperatures $T_{AFE}$ with $r_A$ (reprinted from Ref.~\onlinecite{bcto}). The straight vertical lines in main frame and inset indicate a tolerance factor $t$ equal to 1, obtained for a virtual $A^{2+}$ cation of 1.503 {\AA}. A coordination of 12 was considered for $A^{2+}$, and 6 for Co$^{2+}$, Te$^{6+}$, and O$^{2-}$ in the estimation of $t$; ionic radii extracted from Ref.~\onlinecite{shannon}.}
\label{fig-ap-diel}
\end{figure}

\section{RESULTS AND DISCUSSION}

Figure~\ref{fig-intro} shows a typical ordered perovskite structure, being monoclinically distorted. The regular arrangement of Co$^{2+}$ and $M^{6+}$ cations on the ($B$,$B'$)-sites of the structure is easily visualized. This cation ordering is facilitated by the difference in valence (and size) between the $B$ and $B'$ cations. As can be seen in the left side of Fig.\ref{fig-intro}, the $A^{2+}$ cations on the $A$-site are much larger than $B$- and $B'$-site cations; Co$^{2+}$ on the $B$ site being significantly larger than the typical $M^{6+}$ such as Te$^{6+}$ included in Fig.\ref{fig-intro}.

\begin{figure}[htb]
\includegraphics[width=0.46\textwidth]{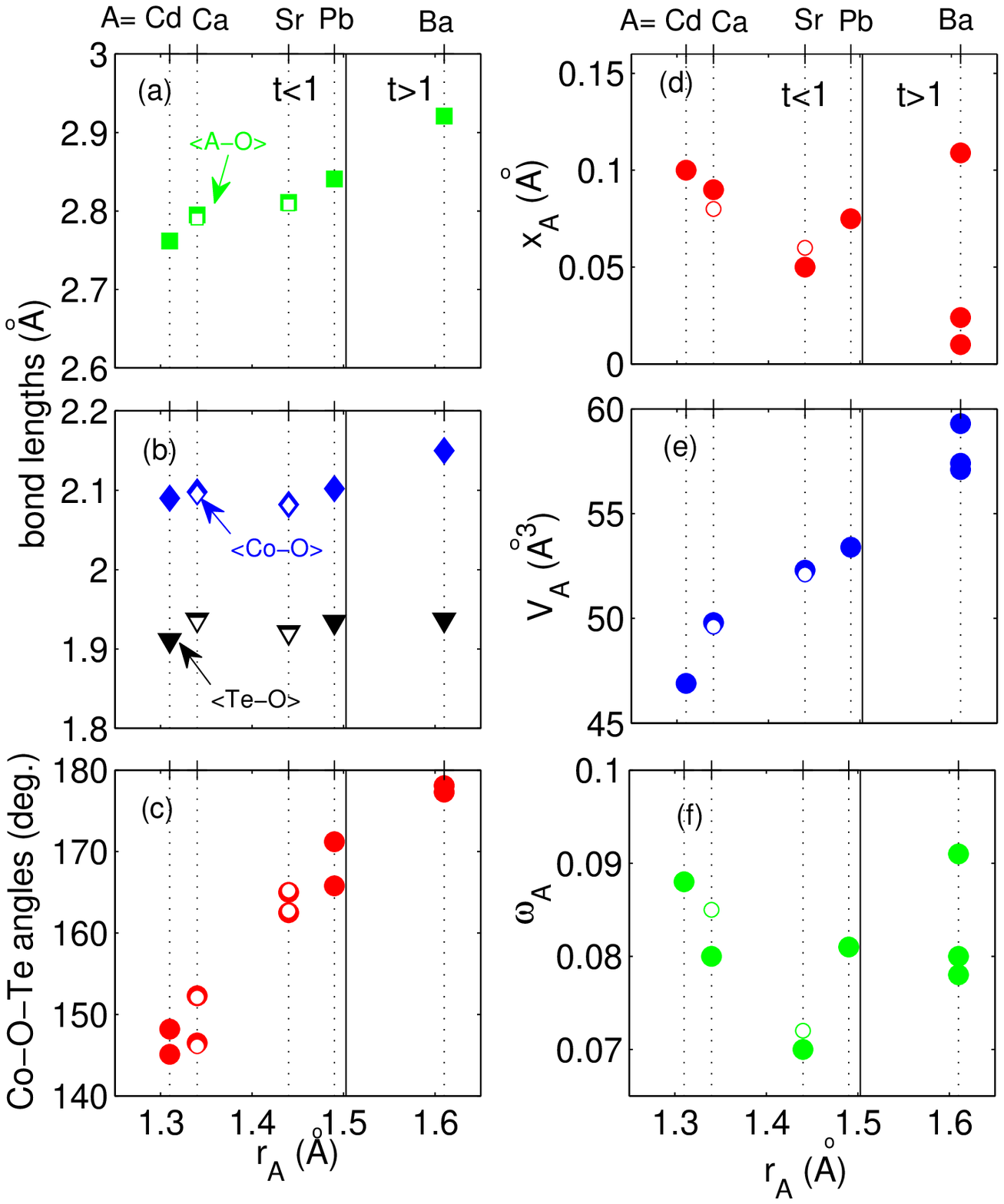}
\caption{(Color online) Dependence on the $A$-cation ionic radius $r_A$ of the low-temperature (T $<$ 10K) structural properties of $A_2$CoTeO$_6$: Left: (a) and (b): average $<A$-O$>$, $<$Co-O$>$, and $<$Te-O$>$ bond lengths, and (c) Co-O-Te bond angles. Right: (d) Shift from centroid $x_A$ (upper panel), (e) $A$O$_{12}$ polyhedral volume $V_A$, and (f) polyhedral volume distortion $\omega_A$ for $A$ cation. Open symbols represent data from Ref.~\onlinecite{alonso}. The straight vertical lines indicate a tolerance factor $t$ equal to 1. The structural data was obtained from XRD measurements ($A$ = Cd, Ca, and Sr) at $T$ = 10 K and NPD measurements at $T$ = 5 K ($A$ = Pb and Ba) and $T$ = 2 K ($A$ = Ca  and Sr from Ref.~\onlinecite{alonso}).} 
\label{fig-struct}
\end{figure}

The tolerance factor $t$ defined as:

$$ t = \frac{r_A+r_O}{\sqrt{2}(\frac{r_B}{2}+\frac{r_{B'}}{2}+r_O)} $$

\noindent where  $r_A$, r$_B$, r$_{B'}$, and $r_O$ are the ionic radii of the different cations and oxygen, describes the stability of the perovskite structure\cite{gold}; $t$ = 1 represents the ideal cubic structure. The compounds with $A$ = Cd, Ca, Sr, and Pb adopt a monoclinic structure $P2_1/n$ at low temperatures, while in order to accommodate the larger Ba cation, Ba$_2$CoTeO$_6$ adopts a $P\overline{3}m$ hexagonal structure. The structure of Ba$_2$CoTeO$_6$ is stable from room-temperature down to low temperatures\cite{bcto}. In the case of Pb$_2$CoTeO$_6$, the tolerance factor is very close to 1 (see e.g. Fig.~\ref{fig-ap-diel}) and the material is cubic above room temperature\cite{pcto}. Pb$_2$CoTeO$_6$ however undergoes successive structural phase transitions as a function of temperature, from cubic $Fm\overline{3}m$ to rhombohedral $R\overline{3}$ (370 K) to monoclinic $I2/m$ (210 K), to monoclinic $P2_1/n$ (125 K), the monoclinic $P2_1/n$ phase being stable down to 5 K. Similar successive transitions were reported for Sr$_2$CoTeO$_6$ as a function of temperature\cite{alonso}, albeit with a direct transition from cubic $Fm\overline{3}m$ to monoclinic $I2/m$ at 773 K. 

\begin{figure}[htb]
\includegraphics[width=0.46\textwidth]{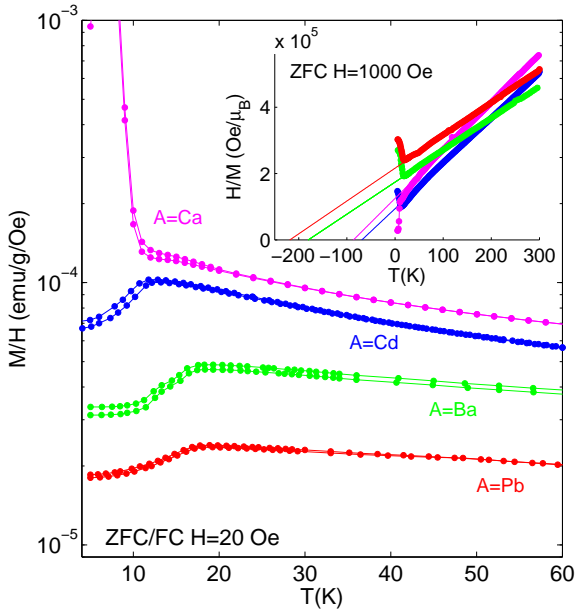}
\caption{(Color online) Temperature $T$ dependence of the zero-field cooled (ZFC) and field-cooled (FC) susceptibility $M/H$ recorded in a small magnetic field $H$ = 20 Oe for $A_2$CoTeO$_6$. The inset shows the ZFC $M/H$ data recorded in a larger field ($H$ = 1000 Oe) plotted as $H/M$, as well as linear fits of the $T$ $>$ 50 K data.}
\label{fig-mt}
\end{figure}

Together with the tolerance factor, one can also define the pseudo-cubic unit cell parameter $a_p$ in order to compare different types of distorted structures:

$$ a_p = (\frac{V}{Z})^{\frac{1}{3}} $$

\noindent where $V$ is the cell volume and $Z$ the number of formula units per cell. The variation of $a_p$ with $r_A$ is shown in Fig.~\ref{fig-ap-diel}. $a_p$ monotonously increases with $r_A$, so that we can use $r_A$ as a representative variable to study the evolution of the structural and magnetic properties for different $A$ cations. For example in the inset of Fig.~\ref{fig-ap-diel}, it can be seen that the antiferroelectric transition temperature $T_{AFE }$ of the compounds monotonously decreases as $r_A$ increases (dielectric data from Ref.~\onlinecite{sergey} for $A$ = Cd, Ca, and Sr, and Refs.~\onlinecite{bcto,pcto} for $A$ = Pb and Ba). Ideally $T_{AFE }$ should be plotted against the shift of $A$ cation from polyhedral centroid position, but this data is not available above room temperature, i.e. near the dielectric anomalies. Two dielectric anomalies are observed as a function of temperature for Pb$_2$CoTeO$_6$ \cite{pcto}, while a relaxor behavior was observed for Ba$_2$CoTeO$_6$ \cite{bcto}, with tolerance factor $t$ $>$ 1.

Figure~\ref{fig-struct} presents the variation of several low-temperature structural quantities with the ionic radius of the $A$ cation. It can be seen that the $<A$-O$>$ bond length follows the increase of $r_A$ from $A$ = Cd to $A$ = Ba. The $<$Co-O$>$ and $<$Te-O$>$ are instead varying much less, especially if considering only the monoclinic ($t$ $<$1) systems. The $<$Co-O-Te$>$ bond angles on the other hand greatly vary with $r_A$, increasing from about 145 ($A$ = Cd) to 177 ($A$ = Ba) degrees. Information on the $A$O$_{12}$,  CoO$_{6}$,  TeO$_{6}$  polyhedra was also extracted\cite{ivton} from the diffraction experiments (cation shift from centroid $x$, polyhedral volume $V$, and polyhedral volume distortion $\omega$). Except for Ba$_2$CoTeO$_6$, the largest variation in the cation shift $x$ and polyhedral distortion $\omega$ were recorded around the $A$ cation. We have thus considered in Fig.~\ref{fig-struct} only the quantities related to the $A$O$_{12}$ polyhedron ($x_A$, $w_A$, and $V_A$). While some variation of the polyhedral shifts and volumes (decrease and increase with increasing $r_A$ respectively) can be seen, the polyhedral distortion is not varying monotonically. The observed increase of $x_A$ with decreasing ionic size of $A$ cation from $A$ = Pb to Cd reflects the increase in distortion as the tolerance factor decreases from values near 1.\\

\begin{figure}[htb]
\includegraphics[width=0.46\textwidth]{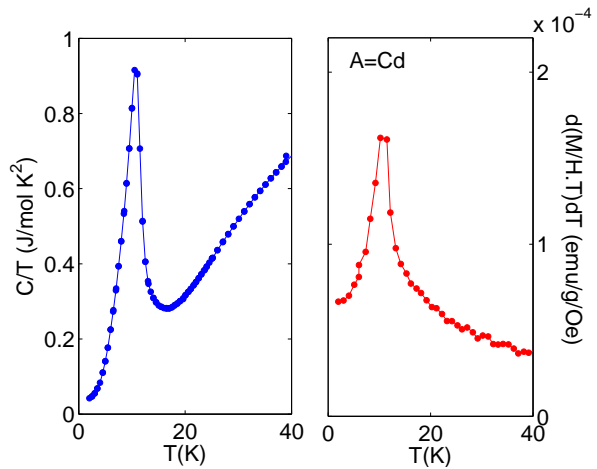}
\caption{(Color online) Temperature dependence of (left) the heat capacity $C$ (plotted as $C/T$) and (right) the temperature derivative of the product $M/H$ $\times$ $T$ for Cd$_2$CoTeO$_6$ (i.e. $A$ = Cd).}
\label{fig-cd}
\end{figure}

We now investigate the magnetic properties of the system. The low-temperature antiferromagnetic ordering can be appreciated from the zero-field-cooled/field cooled (ZFC/FC) magnetization curves presented in the main frame of Fig.~\ref{fig-mt} and heat capacity data (see the example for $A$ = Cd in Fig.~\ref{fig-cd} or data published in Refs.~\onlinecite{bcto,pcto}). Little or no irreversibility is observed between the ZFC/FC curves at any temperature. A clear maximum is observed below 20K for all samples, except for Ca for which a canting-like feature yields a large excess magnetization at the transition.

The derivative $d(M/H \times T)/dT$, which mimics the temperature dependence of the magnetic heat capacity as illustrated in Fig.~\ref{fig-cd}, suggests antiferromagnetic transition temperatures of 10.5 K for $A$ = Cd, 7 K for $A$ = Ca, and 16 K for $A$ = Pb and Ba. These results are in agreement with the previously published data from Ref.~\onlinecite{alonso}, which reported $T_N$ = 10 K for $A$ = Ca and 15 K for $A$ = Sr. Here we should mention that our materials with $A$ = Cd and Ca present some degree (about 20 \%, as estimated from XRD results) of antisite disorder. Such cation disorder is likely to influence the antiferromagnetic interaction, which may explain the somewhat smaller $T_N$ for our Ca$_2$CoTeO$_6$ with respect to the one studied in Ref.~\onlinecite{alonso}, which should be free from such antisite disorder. The difference is only 3K and gives a measure of the potential variation of $T_N$ due to antisite disorder. 

\begin{figure}[htb]
\includegraphics[width=0.46\textwidth]{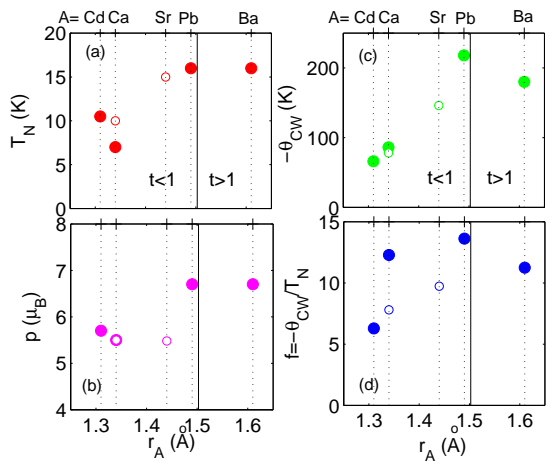}
\caption{(Color online) Dependence on the $A$-cation ionic radius $r_A$ of the magnetic properties of $A_2$CoTeO$_6$: (a) the antiferromagnetic transition temperature $T_N$, (b) effective magnetic moment $p$,  (c) Curie-Weiss temperature $\theta_{CW}$, and (d) the frustration parameter $f$. Open symbols represent data from Ref.\onlinecite{alonso}. The straight vertical lines indicate a tolerance factor $t$ equal to 1.}
\label{fig-allmag}
\end{figure}

Pb$_2$CoTeO$_6$ and Ba$_2$CoTeO$_6$  are free from antisite disorder. In the case of Pb$_2$CoTeO$_6$, a tiny antiferromagnetic-like anomaly (not clearly seen in Fig.~\ref{fig-mt}, see Ref.~\onlinecite{pcto}) is observed below 48 K in the ZFC/FC curves. This anomaly seems intrinsic as it is also observed in single-crystals of Pb$_2$CoTeO$_6$ that we have recently investigated; $T_N$ is also the same as for the ceramic sample. We believe that it is unlikely that this 48 K anomaly is related to e.g. oxygen defficiencies\cite{oxygen} as XRD and NPD (and magnetization, which does not include any ferromagnetic-like phases as observed in Ref.~\onlinecite{oxygen}) results suggest that all of our ceramic samples are rather stoichiometric in cation and oxygen contents. 

Zero-field-cooled/field-cooled curves were also recorded in larger magnetic fields (1000 Oe) in order to investigate the Curie-Weiss behavior of the high-temperature data. A good fit is obtained for the inverse susceptibility $H/M$ (a temperature-independent constant was substracted from the susceptibility) down to just above $T_N$, as shown in the inset of Fig.~\ref{fig-mt}. The obtained variation of the Curie-Weiss parameters (Curie-Weiss temperature $\theta_{CW}$ and effective moment $p$) with the $A$ cation ionic radius is shown in Fig.~\ref{fig-allmag}, together with the variation of $T_N$.

\begin{figure}[htb]
\includegraphics[width=0.46\textwidth]{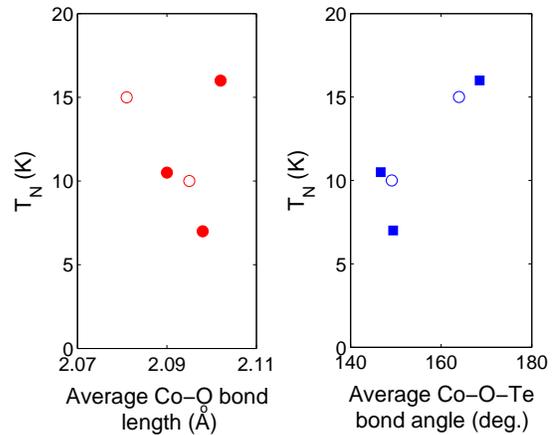}
\caption{(Color online) Dependence of the antiferromagnetic transition temperature $T_N$ on the (a) average $<$Co-O$>$ bond length and (b) average $<$Co-O-Te$>$ bond angle for the low-temperature monoclinic samples ($t$ $<$ 1, i.e. for all but Ba). Open symbols represent data from Ref.~\onlinecite{alonso}.}
\label{fig-tnstruct}
\end{figure}

The antiferromagnetic transition slightly increases as $r_A$ increases. The Curie-Weiss temperature shows a relatively larger variation. The Curie-Weiss temperature is about 6 times larger than $T_N$ for $A$ = Cd (as shown in the lower right panel of Fig.~\ref{fig-allmag} using the frustration parameter $f=-\theta_{CW}/T_N$\cite{frust}), and increases to about 14 times $T_N$ in the $A$ = Pb case. This implies a large magnetic frustration\cite{frust}, which increases with increasing $r_A$. The effective moment seems to slightly increase from about 5.5 $\mu_B$ for $A$ = Cd, Ca and Sr, to about 6.7 $\mu_B$ for $A$ = Pb and Ba compounds. These later values are closer to those of  free Co$^{2+}$ in a high-spin 3$d^7$ configuration, (S = 3/2, L = 3, p = 6.54 $\mu_B$), suggesting a weaker Co-O covalency, possibly related to a simultaneous increase in $A$-O covalency for the Pb and Ba cations. The data obtained in Ref.~\onlinecite{alonso} for $A$ = Ca is nearly identical to ours; in Ref.~\onlinecite{alonso}, the obtained $p$ values were instead compared to the effective moment associated with Co$^{2+}$ in highly regular octahedral environment (5.2 $\mu_B$,  $^4T_{1g}$ ground state).

We have plotted $T_N$ as a function of the average $<$Co-O$>$ bond length and average $<$Co-O-Te$>$ bond angle for the low-temperature monoclinic samples (all but Ba) in Fig.~\ref{fig-tnstruct}. Although it appears to follow the $<$Co-O-Te$>$ bond angle, $T_N$ is only slightly varying, possibly because of the relative weak variation of the $<$Co-O$>$ bond distances.\\

\begin{figure}[htb]
\includegraphics[width=0.46\textwidth]{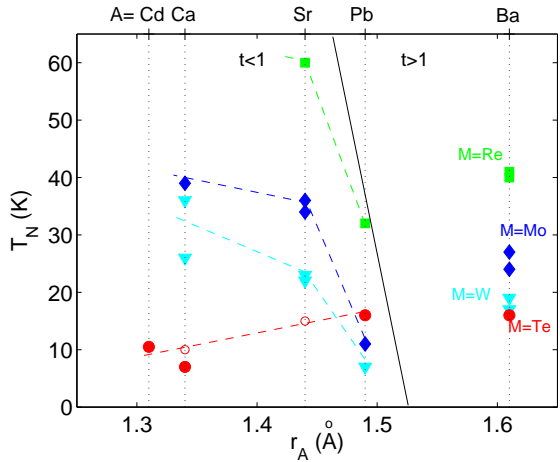}
\caption{(Color online) Dependence on the $A$-cation ionic radius $r_A$ of the antiferromagnetic transition temperature $T_N$ of $A_2$CoTeO$_6$ (large circles). Open symbols represent data from Ref.~\onlinecite{alonso}. Literature data on various $A_2$Co$M$O$_6$ systems is added for comparison, with $M$ = Re (squares), Mo (diamonds), and W (downward triangles). See Refs.~\onlinecite{desc,BaCoMO,PbCoMO,SrCoMO,CaCoMO} for references. Dashed lines are guides to the eye; the straight line indicates a tolerance factor $t$ equal to 1 (the line is not vertical as $t$ depends on the size of $M^{6+}$; $t$ amounts e.g. to 0.998 for $M$ = Re, and decreases to 0.986 for $M$ = W.}
\label{fig-diagall}
\end{figure}

In Fig.~\ref{fig-diagall}, we have plotted the antiferromagnetic transition of $A_2$CoTeO$_6$ as a function of the ionic radius of the $A$ cation, together with literature data on $A_2$Co$M$O$_6$ with $M^{6+}$ = Re$^{6+}$ (5$d^1$), Mo$^{6+}$ (5$d^0$), and W$^{6+}$ (4$d^0$); see Fig.~\ref{fig-intro} for respective ionic sizes (Te$^{6+}$ has $4d^{10}$ configuration). It can immediately be seen, if one considers only the $t$ $<$ 1 region, that the $T_N$ of $A_2$Co$M$O$_6$ compounds with $M$ = Te seems to follow a trend opposite to that of compounds with other $M$ cations; in all the other cases, $T_N$ appears to be more dependent on the size of $A$ cation, increasing with decreasing $A$ cation size. Since all the $M$ cations have similar sizes (Re$^{6+}$ is 0.55 {\AA}, which is close to the 0.56 {\AA} of Te$^{6+}$), this suggest that the above observation is not just a structural effect, but that the Te cations influence the electronic and in turn the magnetic structure of the system differently.  We are not aware of detailed experimental or theoretical studies discussing the specifics of Te$^{6+}$ ions. Yet several groups have reported the more covalent nature of the Te-O bond in perovskites\cite{blasse,woodward2}. Politova and Venevtsev have also reported the vibronic properties and rattling of Te ions in similar oxygen octahedra\cite{politova}, and associated it to the enhanced ferroelectricity of Te-based materials and effect of Te on the electronic properties. In the related Sr$_3$Fe$_2$(Te,W)O$_9$ perovskite, the decrease of $T_N$ upon substitution of W for Te was similarly associated to the difference in electronic states of $W^{6+}$ and $Te^{6+}$; see Ref.~\onlinecite{SFWTO} and references therein. 

Another interesting observation is that for the compounds with $A$ = Pb, closest to $t$ = 1, the $T_N$ of the Pb$_2$CoMO$_6$ serie monotonously decreases as the ionic size of the $M$ cation increases (see Fig.~\ref{fig-diagall}). Such monotonous variation is not observed for the other $A$ cations in the $t$$<$ 1 region, nor in the  $t$ $>$ 1 region.  Two members of the $A_2$CoUO$_6$ serie ($M$ = U, ionic radius $r_U$ = 0.73 {\AA}) have been synthesized, with $A$ = Sr ($T_N$ = 10 K) and Ba ($T_N$ = 9K) \cite{SrCoUO,BaCoUO}; hence unfortunately only one system in the $t$ $<$ 1 region.\\

The electronic phase diagram displayed in  Fig.~\ref{fig-diagall} suggests that compounds such as Cd$_2$CoReO$_6$ or Ca$_2$CoReO$_6$ may exhibit relatively large $T_N$, making them more attractive for potential application. Interestingly Ca$_2$CoReO$_6$ was reported to have a conical spin structure with associated excess moment below 130 K\cite{tokura-Re}, while the related compound Ca$_2$FeReO$_6$ was found to order ferrimagnetically at high temperatures (540 K), displaying large electronic correlation effects\cite{tokura-dd}. Note that in that case Re might exist in other valence states as Re$^{6+}$. This indicates that $A_2B$ReO$_6$ ($B$ = Co, Ni, Fe, etc) perovskites may be an interesting base to produce new multiferroics.\\

\section{CONCLUSIONS}

To summarize, we have investigated the structural and magnetic properties of antiferroelectric $A_2$CoTeO$_6$ perovskites with $A$ = Cd, Ca, Sr, Pb, and Ba. All compounds were found to be antiferromagnetic at low temperatures with N\'eel temperatures slightly decreasing with the decreasing ionic size of the $A$ cation. Such a decrease in $T_N$ is not observed in related $A_2$Co$M$O$_6$ materials with other $M^{6+}$ cations of similar size, suggesting an electronic effect of the Te$^{6+}$ ions. Interestingly, a monotonous variation of $T_N$ with the ionic size of $M$ cation was observed in Pb$_2$CoMO$_6$ (tolerance factor $t$ near 1). 

\begin{acknowledgments}
We thank the Swedish Research Council (VR), the G\"oran Gustafsson Foundation, the Royal Swedish Academy of Sciences, and the Russian Foundation for Basic Research for financial support. We are grateful to J. A. Alonso for useful discussions and for providing the structural parameters presented in Ref.~\onlinecite{alonso}, and to M. Weil for providing the single crystals of Pb$_2$CoTeO$_6$.
\end{acknowledgments}

\end{document}